\documentclass[12pt]{iopart}
\usepackage{epsfig} 
\usepackage{bm} 

\begin{document}
\title{Entanglement at finite temperatures in the electronic two-particle interferometer} 
\author{P. Samuelsson$^1$, I. Neder$^2$,
M. B\"uttiker$^3$} 
\address{$^1$Division of Mathematical Physics,
Lund University, Box 118, S-221 00 Lund, Sweden}
\address{$^2$Physics Department, Harvard University, Cambridge,
Massachusetts 02138, USA} 
\address{$^3$D\'epartement de Physique
Th\'eorique, Universit\'e de Gen\`eve, CH-1211 Gen\`eve 4,
Switzerland}
 
\begin{abstract} 
In this work we discuss a theory for entanglement generation,
characterization and detection in fermionic two-particle
interferometers at finite temperature. The motivation for our work is
provided by the recent experiment by the Heiblum group, Neder {\it et
al}, Nature {\bf 448}, 333 (2007), realizing the two particle
interferometer proposed by Samuelsson, Sukhorukov, and B\"uttiker,
Phys. Rev. Lett. {\bf 92}, 026805 (2004). The experiment displayed a
clear two-particle Aharonov-Bohm effect, however with an amplitude
suppressed due to finite temperature and dephasing. This raised
qualitative as well quantitative questions about entanglement
production and detection in mesoscopic conductors at finite
temperature. As a response to these questions, in our recent work,
Samuelsson, Neder, and B\"uttiker, Phys. Rev. Lett. {\bf 102}, 106804
(2009) we presented a general theory for finite temperature
entanglement in mesoscopic conductors. Applied to the two-particle
interferometer we showed that the emitted two-particle state in the
experiment was clearly entangled. Moreover, we demonstrated that the
entanglement of the reduced two-particle state, reconstructed from
measurements of average currents and current cross correlations,
constitutes a lower bound to the entanglement of the emitted
state. The present work provides an extended and more detailed
discussion of these findings.
\end{abstract}

\maketitle

\section{Introduction}

There is presently a strong interest in computation and information
processing based on fundamental principles of quantum mechanics
\cite{Nielsen}. Quantum information technology has the potential both
to address problems that can not be solved by standard, classical
information technology as well as to radically improve the performance
of existing classical schemes. The prospect of scalability and
integrability with conventional electronics makes solid state systems
a likely future arena for quantum information processing. Of
particular interest is the entanglement between the elementary charge
carriers, quasiparticles, in meso- or nanoscopic solid state
conductors. Entanglement, or quantum mechanical correlations,
constitutes a resource for any quantum information process. Moreover,
due to controllable system properties and coherent transport
conditions, conductors on the meso and nano scale constitute ideal
systems for the generation and detection of quasiparticle
entanglement. This opens up for quantum bits based on the spin or
orbital quantum states of individual electrons, the ultimate building
blocks for solid state quantum information processing.

To date quasiparticle entanglement has however remained experimentally
elusive. In particular, there is no unambiguous experimental
demonstration of entanglement between two spatially separated
quasiparticles. A class of mesoscopic systems that appear promising
for a successful entanglement experiment are conductors without direct
interactions between the quasiparticles. It was shown by Beenakker
{\it et al} \cite{Been03} that fermions emitted from a thermal source
can, in contrast to bosons, be entangled by scattering at a
beam-splitter. This was originally discussed for electron-hole pairs
\cite{Been03} and shortly afterward for pairs of electrons
\cite{Sam04,Been04a}. Since then there has been a large number of
works on entanglement of non-interacting particles, see
e.g. \cite{nonint1,nonint2,nonint3,Titov,Kindermann,Frustaglia} for a number
of representative papers and also \cite{Beenrev} for a review.

Several of the entanglement proposals have been based on electrical
analogs of optical interferometers and beam-splitter geometries. Such
electronic systems are conveniently implemented in conductors in the
quantum Hall regime, where electrons propagate along chiral edge
states \cite{halp,mb88} and quantum point contacts constitute
reflectionless beam-splitters \cite{BS1,BS2,BS3} with controllable
transparency, see e.g. \cite{Buttpap}. Recent experimental progress on
electronic Mach-Zehnder \cite{MZ1,MZ2,MZ3,MZ4,MZ5} and Hanbury Brown
Twiss \cite{Neder} interferometers has provided further motivation for
a theoretical investigation of entanglement in such systems. In
addition, the experimental realization \cite{Feve} of time-controlled
single-electron emitters \cite{singem1,singem2} in quantum Hall
systems has opened up the possibility for a dynamical generation of
entangled quasiparticles, entanglement on demand
\cite{Timeent1,Timeent2,Timeent3,Timeent4}.

In this work we will focus on the electronic two-particle, or Hanbury
Brown Twiss, interferometer. A theoretical proposal for an
implementation of this two-particle interferometer (2PI) in a
conductor in the quantum Hall regime was proposed by two of us, P.S
and M.B., together with E. V. Sukhorukov in Ref. \cite{Sam04}. Recently,
the Heiblum group, including one of us, I.N., was able to realize the
2PI in a versatile system which could be electrically tuned between
with two independent Mach-Zehnder interferometers and a 2PI. In
perfect agreement with the theoretical predictions \cite{Sam04}, the
two-particle interference pattern was visible in the current
correlations but not in the average current. As discussed in
Ref. \cite{Sam04}, there is an intimate relation between two-particle
interference and entanglement in the fermionic 2PI. Under ideal
conditions, i.e. zero temperature and perfect coherence, two-particle
interference implies that the two particle wave function is on the
form
\begin{equation}
|\Psi_s\rangle=\frac{1}{\sqrt{2}}\left[|1\rangle_A|2\rangle_B-|2\rangle_A|1\rangle_B\right].
\label{introsing}
\end{equation}
Here $1,2$ denote the sources and $A,B$ the sites of detection, as
shown in Fig. \ref{system}.  The wavefunction $|\Psi_s\rangle$ is
maximally entangled, it is a singlet in the orbital, or pseudo spin,
space $\{|1\rangle, |2\rangle \}$.

However, in the experiment \cite{Neder}, $\sim 25\%$ visibility of the
current correlation oscillations was observed. This indicates that
both decoherence and finite temperature is important. Dephasing can
qualitatively be accounted for \cite{Sam03,Turkbeen,Turksam} by a
suppression of the off-diagonal components of the density matrix
$|\Psi_s\rangle\langle \Psi_s|$. It was shown that at zero temperature the
entanglement survives for arbitrary strong dephasing. The effect of
finite temperature was not investigated at the time of the experiment.

The experimental findings thus raised two important questions: are the
electrons reaching the detectors at A and B entangled and if so, can
this two-particle entanglement be unambiguously detected by
measurements of currents and current correlators, the standard
quantities accessible in electronic transport measurements? In our
recent work \cite{Sam09} we provided a positive answer to both these
questions. We first calculated the entanglement of the emitted
two-particle state and found that the state was clearly
entangled. Thereafter we showed that under very general conditions the
entanglement of the reduced two-particle density matrix provides a
lower bound for the entanglement of the emitted two-particle
state. Since the reduced density matrix is possible to reconstruct
tomographically by current and current correlation measurements
\cite{tomo}, this provides an unambiguous way to detect the
entanglement of the emitted state. In the present paper we discuss
these findings in more detail.

\section{The two-particle interferometer in optics and electronics}

\begin{figure}[h]
\centerline{\psfig{figure=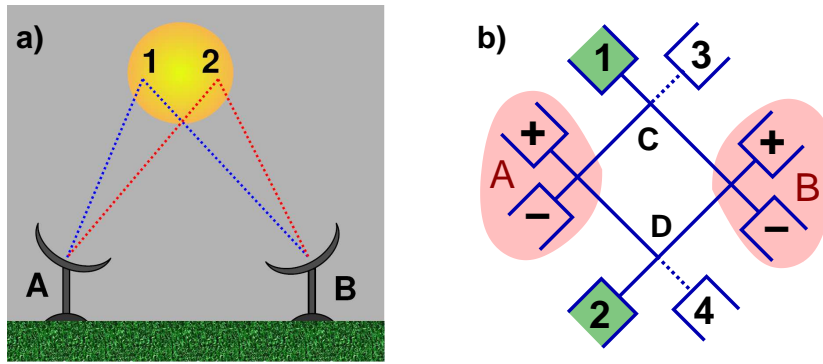,width=11.0cm}}
\caption{a) Schematic of the Hanbury Brown Twiss intensity
interferometer used to measure the angular diameter of stars. Two
uncorrelated points 1,2 on the star act as sources. The signal is
detected at A and B. b) Schematic of the topologically equivalent
two-particle interferometer (2PI) \cite{Sam04} with beam splitters C,D
and biased, active (grounded, inactive) source contacts 1,2
(3,4). Detector regions A and B (red shaded) contain beam splitters
and grounded contacts $\pm$.}
\label{system}
\end{figure}

Interference is most often investigated in structures that lead to a
superposition of amplitudes of a single particle. However, in 1956,
Hanbury Brown and Twiss (HBT) invented an optical interferometer based
on correlations of light intensities \cite{HBT1,HBT2}, an optical 2PI, see
fig. \ref{system}. The intensity interferometer allowed HBT to
determine the angular diameter of a number of visual stars, not
possible with available single particle, or Michelson,
interferometers.  The HBT intensity interferometer displays two
distinct but fundamentally interrelated features: \\

$\bullet$ First, there is a direct statistical effect since photons
from a thermal light source tend to bunch, whereas fermions would
anti-bunch. This effect has been used in a large number of experiments
in different fields of physics such as elementary particles
\cite{Baym}, solid state \cite{BS1,BS2,BS3} and free \cite{vacuum}
electrons and recently cold atoms \cite{Coldat}.\\

$\bullet$ Second, light from two different, completely uncorrelated
sources gives rise to an interference effect in intensity correlations
but not in the intensities themselves. This is the two-particle
interference effect. In optics, various aspects of two-particle
interference have been investigated extensively since the
HBT-experiment, see e.g. \cite{Mandel} for a short review, and is
still a subject of interest \cite{Zeil}. In electronics, only very
recently was a fermionic two-particle interferometer realized
\cite{Neder}, the subject of this work. \\

Fundamentally both of these effects are related to the symmetry of the
multiparticle wave function under exchange of two particles. We note
that albeit the HBT-experiment could be explained by a classical
electro-magnetic theory, a compelling quantum mechanical picture based
on individual photons was put forth soon after the experiment
\cite{Purcell}. Importantly, for fermions no classical theory exists.

To obtain a qualitative understanding of the physics of two-particle
interferometers it is rewarding to compare the properties of optical,
bosonic interferometers and electronic, fermionic interferometers.  In
Fig. \ref{system} a schematic of a two-particle interferometer,
topologically equivalent to the HBT-interferometer, is shown. A
natural measure of the correlations between the particles at $A$ and
$B$ is the probability to jointly detect one particle at $A$ and one
at $B$. An expression for this joint detection probability for photons
was derived by Glauber \cite{Glauber}. In Ref. \cite{Sam04} this was
adapted to detection of electrons. Here we consider the probability to
detect one photon/electron in detector $A\alpha$, $\alpha=\pm$, at
time $t$ and one in detector $B\beta$, $\beta=\pm$ at a time $t+\tau$,
given by
\begin{eqnarray}
P_{A\alpha B\beta}(\tau)\propto\langle
b^{\dagger}_{B\beta}(t)b^{\dagger}_{A\alpha}(t+\tau)b_{A\alpha}(t+\tau)b_{B\beta}(t)\rangle
\label{jdp}
\end{eqnarray}
The photon/electron creation operators at A are
$b^{\dagger}_{A\alpha}(t)=\int
dE~\mbox{exp}(iEt/\hbar)b_{A\alpha}^{\dagger}(E)$, with
$b_{A\alpha}^{\dagger}(E)$ creating a particle in $A\alpha$ at energy
$E$ and similarly at B. For photons we consider thermal sources in
$1$ and $2$ while $3$ and $4$ are left empty. A detector frequency
window of size $\Delta \omega$ is assumed, over which the distribution
functions of the sources are constant, i.e. $\Delta \omega \ll
kT$. For electrons we assume zero temperature and the sources $1$ and
$2$ biased at $eV$ while sources $3$ and $4$ are grounded. Only
quasiparticle excitations, $E \geq 0$ are considered.

The probabilities are normalized such that $\sum_{\alpha,\beta=\pm}
P_{A\alpha B\beta}=1$. Following the scattering theory for
intensity/current correlations for bosons/fermions emitted from
thermal sources \cite{mb90,mb92}, we get
\begin{eqnarray}
P_{A\alpha B\beta}(\tau)&\propto
&|s_{A\alpha 1}|^2|s_{B\beta 1}|^2\left[1\pm
g(\tau)\right]+|s_{A\alpha 2}|^2|s_{B\beta 2}|^2\left[1\pm g(\tau)\right]
\nonumber \\
&+&|s_{A\alpha 1}|^2|s_{B\beta 2}|^2+|s_{A\alpha 2}|^2|s_{B\beta 1}|^2 \nonumber \\
&\pm &
g(\tau)\left[s_{A\alpha 1}^*s_{B\beta 2}^*s_{B\beta 1}s_{A\alpha 2}+s_{A\alpha 1}s_{B\beta 2}s_{B\beta 1}^*
s_{A\alpha 2}^*\right]
\label{jdpt}
\end{eqnarray}
where $g(\tau)=\sin^2(\tau/\pi\tau_C)/(\tau/\pi\tau_C)^2$ contains the
time dependence, with $\tau_C=h/eV$ the coherence time for electrons
and $2/\pi \Delta\omega$ for photons. Here $s_{A\alpha 2}$ is the amplitude
to scatter from source 2 to detector $A\alpha$ etc. The upper/lower signs $\pm$
correspond to electrons/photons.

Several interesting conclusions can be drawn directly from
Eq. (\ref{jdpt}):
 
1) For $\tau \gg \tau_C$, $g(\tau)$ approaches zero and
$P_{A\alpha B\beta}$ is just proportional to the product of the two mean
currents/intensities. The fermionic versus bosonic statistics of the
particle plays no role.

2) For shorter times, $\tau \leq \tau_C$, $g(\tau)$ is finite and the
statistics is important. Note that, as pointed out above, that the
statistics of the particles enter in two different ways.\\ i) The
first two terms in Eq. (\ref{jdpt}) describe a direct bunching (+) or
anti-bunching (-) effect for two particles emitted from the same
reservoir within a time $\tau \leq \tau_C$. This effect would still be
present if one of the sources $1$ or $2$ is removed. \\ ii) The last
two terms describe the two-particle, or exchange \cite{mb90,mb92},
interference, where the $\pm$ sign explicitly follows from the
interchange of the two detected particles. This two particle
interference is only present when both sources are
active.

For semitransparent beam-splitters $A,B,C$ and $D$ and coincident
detection $\tau\ll \tau_C$ we have
\begin{eqnarray}
P_{A\alpha B\beta}=\left\{ \begin{array}{cc} \frac{1}{4}\left[1+\alpha\beta\cos \phi\right] & \mbox{fermions} \\ \frac{1}{4}\left[1+\frac{\alpha\beta}{2}\cos \phi\right] & \mbox{bosons} \end{array}\right.
\label{jdpt2}
\end{eqnarray}
where $\phi$ is a scattering phase. From this expression a very
important difference between bosonic and fermionic thermal sources is
apparent: the visibility
\begin{equation}
\nu=\frac{P_{A\alpha B\beta}^{max}-P_{A\alpha B\beta}^{min}}{P_{A\alpha B\beta}^{max}+P_{A\alpha B\beta}^{min}}
\end{equation}
of the oscillations is $1$ for fermions but only $1/2$ for
bosons. This is directly related to the fact that while the emitted
fermionic two-particle state is maximally entangled, the bosonic state
is unentangled \cite{yurke}.

\section{Fermionic two particle interferometer: theory} 

\begin{figure}[h]
\centerline{\psfig{figure=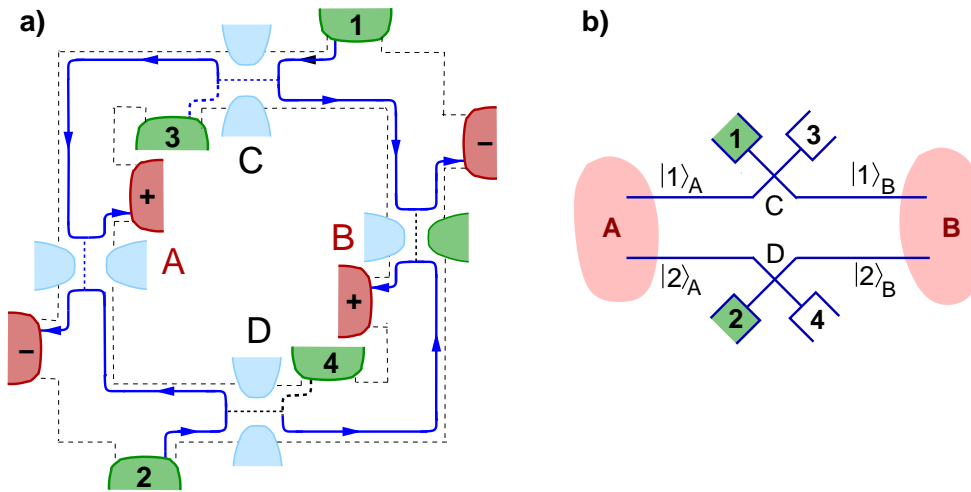,width=13.0cm}}
\caption{a) Fermionic 2PI implemented in a conductor in the quantum
Hall regime, from \cite{Sam04}.  See text for details. b) Schematic of
the source part of the 2PI, with the the orbital states
$|1\rangle_A,|2\rangle_A,|1\rangle_B$ and $|2\rangle_B$ for particles
emitted out from the source towards the detectors are shown.}
\label{HBTferm}
\end{figure}

In Ref. \cite{Sam04} we proposed an implementation of an electronic
2PI in a conductor in the quantum Hall regime, with electrons
propagating along single, spin polarized edge states (see
Fig. \ref{HBTferm}). Two electronic reservoirs $1,2$ biased at $eV$
act as sources for electrons while the reservoirs $3,4$ as well as the
detector reservoirs are grounded. All reservoirs are kept at the same
temperature $T$. Moreover, we consider here only the linear regime in
voltage where electron-electron interactions can be neglected. This
regime is relevant for the experiment \cite{Neder}. The QPC's at
$A,B,C$ and $D$ act as beamsplitters with transparencies $T_A,T_B,T_C$
and $T_D$ respectively.

The scattering amplitude $s_{A+1}=\sqrt{T_AR_C}e^{i\phi_{AC}}$, where
$R_C=1-T_C$ and $\phi_{AC}$ is the scattering phase picked up by the
electron up when traveling from C to A. Similar relations hold for the
other scattering amplitudes. Note that the total phase
$\phi=\phi_{AC}-\phi_{AD}+\phi_{BD}-\phi_{BC} $ is, up to a constant
term, given by $2\pi \Phi/\Phi_0$ where $\Phi$ is the magnetic flux
threading the 2PI and $\Phi_0=h/e$, the single particle flux
quanta. Importantly, the Corbino geometry in Fig. \ref{HBTferm} with
unidirectional edge states and reflectionless beam-splitters is
topologically equivalent to the 2PI shown in Fig. \ref{system}.

\subsection{Two particle Aharonov-Bohm effect} 

The standard tools for investigating transport properties in
mesoscopic electronic systems are average electrical current and
current correlation measurements \cite{Buttrev}. A scattering theory
calculation \cite{mb86} gives the average current at contact $A\alpha$
\begin{eqnarray}
I_{A\alpha} &=&\frac{e}{h}\int dE \left(|s_{A\alpha 1}|^2+|s_{A\alpha 2}|^2\right)\left[f_V(E)-f(E)\right] \nonumber \\
\label{curr}
\end{eqnarray}
and similar at $B \beta$. Here $f_V=1/(1+e^{(E-eV)/kT})$ and
$f=1/(1+e^{E/kT})$ are the Fermi distributions of the biased,
$1,2$ and the grounded, $3,4$ reservoirs respectively. The irreducible
zero frequency correlator 
\begin{equation}
S_{A\alpha B\beta}=\int dt \langle \Delta
I_{A\alpha}(0)\Delta I_{B \beta}(t)\rangle
\end{equation}
between currents $I_{A
\alpha}(t)=I_{A \alpha}+\Delta I_{A \alpha}(t)$ and $I_{B
\beta}(t)=I_{B \beta}+\Delta I_{B \beta}(t)$ \cite{mb92} becomes
\begin{equation}
S_{A\alpha B\beta}=\frac{e^2V}{h}\int dE \left(|s_{A\alpha 1}^*s_{B\beta 1}+s_{A\alpha 2}^*s_{B\beta 2}|^2\right)\left[f_V(E)-f(E)\right]^2
\label{noise}
\end{equation}
These expressions are valid for arbitrary temperature but for the rest
of the discussion in this section we only consider the zero
temperature case. In particular, for the simplest possible case, with
all beam-splitters semitransparent and energy-independent scattering
amplitudes, we have
\begin{eqnarray}
I_{A\alpha}=I_{B\beta}=\frac{e^2V}{2h}, \hspace{0.5cm} S_{A\alpha B\beta}=\frac{e^3V}{4h}\left[1+\alpha\beta\cos\phi\right] 
\label{currnoise}
\end{eqnarray}
While the average current is a function of QPC-transparencies only,
the current cross correlator depends also on the phase $\phi$. Since
this phase is proportional to the magnetic flux $\Phi$ threading the
2PI, we call this a two-particle Aharonov-Bohm (AB) effect.

Interestingly, we can directly relate the coincident detection
probability in Eq. (\ref{jdpt}) at times $\tau\ll\tau_C$ with the
currents in Eq. (\ref{curr}) and the zero frequency noise correlators
in Eq. (\ref{noise}) as $[g(0)=1]$
\begin{eqnarray}
P_{A\alpha B\beta}(0) &\propto& |s_{A\alpha 1}s_{B\beta 1}^*+s_{A\alpha 2} s_{B\beta 2}^*|^2+\left(|s_{A\alpha 1}|^2+|s_{A\alpha 2}|^2\right)\left(|s_{B\beta 1}|^2+|s_{B\beta 2}|^2\right) \nonumber \\
& \propto& S_{A\alpha B\beta}+2\tau_CI_{A\alpha}I_{B \beta}
\label{jdptot}
\end{eqnarray}
This is a direct consequence of fermionic anti-bunching, leading to a
filled stream of electrons emitted from the source reservoirs and
hence making long time observables an effective average of many
individual, short time, single and two-particle events.

\subsection{Entanglement}

The connection between this two-particle Aharonov-Bohm effect and
entanglement can be seen by considering the many-body ground state
$|\Psi_{in}\rangle$ of the electrons injected into the 2PI. Electrons
at different energies are independent and the many-body state at zero
temperature is thus a product state in energy
\begin{equation}
|\Psi_{in}\rangle=\prod_{0\leq E \leq
eV}  a_1^{\dagger}(E)a_2^{\dagger}(E)|\bar 0\rangle
\end{equation}
where $|\bar 0\rangle$ is the filled Fermi sea and $a^{\dagger}_1(E)$
creates an electron at energy $E$, incident from reservoir
$1$. Adopting the formalism of Ref. \cite{Been03} we first define
$|\Psi_{in}(E)\rangle=a_1^{\dagger}(E)a_2^{\dagger}(E)|\bar 0\rangle$
the injected state at energy $E$. We have the scattering relations at
the two source beam splitters, suppressing energy notation
\begin{equation}
\left(\begin{array}{c} b_{A1} \\ b_{B1} \end{array} \right)
=\left(\begin{array}{cc} r_C & t_C' \\ t_C & r_C'
 \end{array} \right) \left(\begin{array}{c} a_{1} \\ a_{3} \end{array} \right),  \hspace{0.5cm}
\left(\begin{array}{c} b_{A2} \\ b_{B2} \end{array} \right)
=\left(\begin{array}{cc} r_D & t_D' \\ t_D &
r_D' \end{array} \right) \left(\begin{array}{c} a_{2} \\ a_{4} \end{array} \
\right)
\label{scatrel}
\end{equation}
for incoming (a's) and outgoing (b's) electrons. The primed scattering
amplitudes thus describes particles incoming from the unbiased
sources. This gives the emitted state for the electrons at energy $E$,
after beam-splitters $C,D$ but before impinging on the detector beam
splitters $A,B$, as
\begin{equation}
|\Psi_{out}(E)\rangle=\left(r_Cb_{A1}^{\dagger}+t_Cb_{B1}^{\dagger}\right)\left(r_Db_{A2}^{\dagger}+t_Db_{B2}^{\dagger}\right)|\bar 0\rangle
\end{equation}
Since we are interested in entanglement between particles in the two,
spatially separated detector regions A and B we project out the part
of the wave function with one particle in A and one in B yielding the
normalized wavefunction
\begin{equation}
|\Psi_{AB}(E)\rangle=\frac{1}{\sqrt
 N}\left(r_Ct_D b_{A1}^{\dagger}b_{B2}^{\dagger}-r_Dt_Cb_{A2}^{\dagger}b_{B1}^{\dagger}\right)|\bar
 0\rangle
\end{equation}
with $N=|r_Dt_C|^2+|r_Ct_D|^2=R_CT_D+R_DT_C$ the normalization
constant. Here we introduced the transmission and reflection
probabilities of the source beam splitters as $T_C=|t_C|^2=|t_C'|^2$ and
$R_C=|r_C|^2=|r_C'|^2=1-T_C$ for C and similarly for D. To make this
more transparent we can, since the two particles live in well
separated Hilbert spaces, introduce the Dirac notation
$|1\rangle_A\equiv b_{A1}^{\dagger}|\bar 0\rangle$ etc, and write
\begin{equation}
|\Psi_{AB}(E)\rangle=\frac{1}{\sqrt{N}}\left[r_Ct_D|1\rangle_A|2\rangle_B-t_Cr_C|2\rangle_A|1\rangle_B\right]
\label{pureAB}
\end{equation}
which for semi-transparent beam splitters (and scattering phase
$\phi=0$) reduces to the singlet state $|\Psi_s\rangle$ in
Eq. (\ref{introsing}). The orbital states are shown in
Fig. \ref{HBTferm}

The entanglement of the state $|\Psi_{AB}(E)\rangle$ can conveniently
be quantified in terms of the concurrence $C$ \cite{Wooters}, which
ranges from zero for an unentangled state to unity for a maximally
entangled state. Working in the computational basis
$\{|1\rangle_A|1\rangle_B,|1\rangle_A|2\rangle_B,|2\rangle_A|1\rangle_B,|2\rangle_A|2\rangle_B\}$,
for the pure state $|\Psi_{AB}\rangle$ in Eq. (\ref{pureAB}) we have
\begin{equation}
C=|\langle \Psi_{AB}|(\sigma_y\otimes \sigma_y)|\Psi_{AB}^*\rangle|
\end{equation}
where $|\Psi_{AB}^*\rangle$ is $|\Psi_{AB}\rangle$ with all
coefficients complex conjugated, $\sigma_y$ a Pauli matrix and
$\otimes$ the direct, tensor product. We thus find for
$|\Psi_{AB}\rangle$ the concurrence
\begin{equation}
C=\frac{2}{N}|r_Ct_Cr_Dt_D|=\frac{2}{N}\sqrt{R_CT_CR_DT_D}
\end{equation}
which reaches unity for semitransparent beam splitters, i.e. for the
singlet state in Eq. (\ref{introsing}). Note that the normalization
factor $N$ is maximal, equal to $1/2$, for semitransparent beam
splitters. This demonstrates that at most only half of the particles
injected from $1$ and $2$ lead to split pairs, with one particle
emitted towards $A$ and one towards $B$, i.e. a maximal pair emission
rate of $1/2$. For a measurement during a time $\tau$ the maximum
concurrence production \cite{Beenrev} is thus ${\mathcal N}/2$, where
${\mathcal N}=\tau eV/h$ the number of pairs injected from $1$ and $2$
in the time $\tau$ and energy interval $0\leq E\leq eV$

\subsection{Dephasing}
There are several microscopic mechanisms that can lead to dephasing,
typically suppressing the two-particle interference. For low
temperatures it is commonly believed that the dominatinating mechanism
for dephasing is electron-electron interactions, but this is still a
topic of ongoing research and goes beyond the scope of the present
work. Here we consider no specific mechanism but model dephasing
qualitatively by coupling one of the interferometer arms to a
dephasing voltage probe \cite{probe1,probe2,probe3,probe4}. In this
context we point out a recent experiment \cite{probeexp}: a voltage
probe was coupled, via a tunable quantum point contact, to one arm of
a Mach Zehnder interferometer in the quantum Hall regime,
demonstrating controllable dephasing. Considering semitransparent beam
splitters, the dephasing probe coupled with a strength $0 \leq \gamma
\leq 1$ lead to a modification of the current correlator in
Eq. (\ref{currnoise}) to \cite{Vanessa}
\begin{eqnarray}
S_{A\alpha B\beta}^{deph}=\frac{e^3V}{4h}\left[1+\gamma\alpha\beta\cos\phi\right] 
\end{eqnarray}
From this expression it is clear that $\gamma$ enters as a decoherence
parameter; decreasing $\gamma$ from $1$ to $0$ leads to a suppression
the phase dependence of the current correlator. In the presence of
dephasing the emitted state is no longer a pure state, it is instead a
mixed state described by a density matrix $\sigma_{AB}$. Considering
zero temperature, working in the computational basis the result for
$S_{A\alpha B\beta}^{deph}$ corresponds to a suppression of the
off-diagonal components of $|\Psi_{AB}\rangle \langle \Psi_{AB}|
\rightarrow \sigma_{AB}$ as
\begin{equation}
\sigma_{AB}=\frac{1}{2}\left(\begin{array}{cccc} 0 & 0 & 0 &0 \\ 0 & 1 & -\gamma & 0 \\ 0 & -\gamma & 1 & 0\\ 0 & 0 & 0 &0 \end{array}\right)
\end{equation}
The concurrence for a mixed state is \cite{Wooters}
\begin{equation}
C=\mbox{max}\left\{\sqrt{\lambda_1}-\sqrt{\lambda_2}-\sqrt{\lambda_3}-\sqrt{\lambda_4},0\right\}
\end{equation}
where $\lambda_i$, $i=1-4$, are the eigenvalues in decreasing order of
$\sigma_{AB}(\sigma_y\otimes \sigma_y) \sigma^*_{AB} (\sigma_y \otimes
\sigma_y)$. We then have
\begin{equation}
C=\gamma
\end{equation}
This means that the entanglement persists even for very strong
dephasing \cite{Sam03, Turkbeen,Turksam}. This is a consequence of the
2PI-geometry, where scattering between the arms, i.e. pseudo spin-flip
scattering, is prohibited.

\subsection{Fermionic two particle interferometer: experiment}

\begin{figure}[h]
\centerline{\psfig{figure=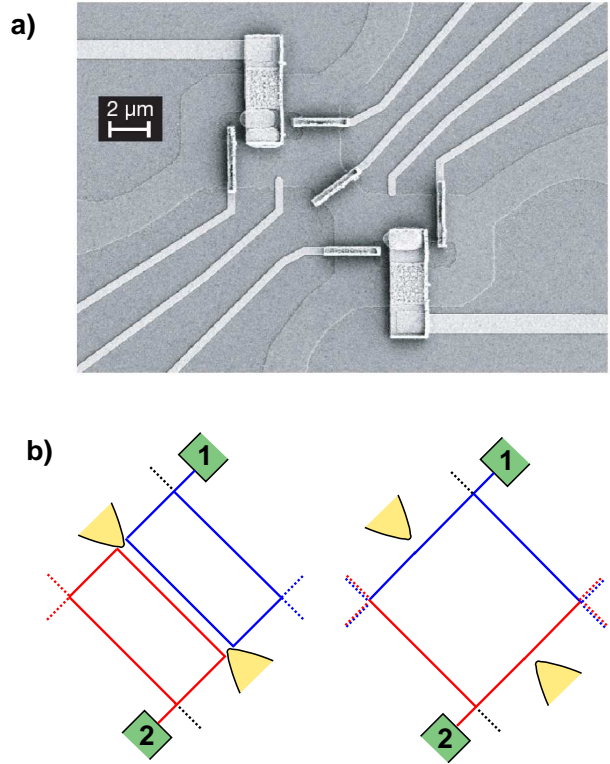,width=8.0cm}} 
\caption{Fermionic two-particle interferometer implemented in a
conductor in the quantum Hall regime in Ref \cite{Neder}. a) Figure
reproduced from Ref. \cite{Neder}. Micrograph of the sample. b) Left:
The system in the two Mach Zehnder interferometers
configuration. Right: The system in the 2PI configuration.}
\label{expfig}
\end{figure}

Very recently the electronic 2PI was realized experimentally by Neder
{\it et al}. In the experiment, in the quantum Hall regime, it was
possible to electrically tune the system between two individual Mach
Zehnder interferometers and a 2PI, as shown schematically in
fig. \ref{expfig}.
The authors first tuned the system to two Mach-Zehnder interferometers
and measured the single particle interference in the average current
for each interferometer. They found a very large visibility in both
interferometers, around $80\%$. They also determined the periods of
the single particle AB-oscillations as a function of both the area and
the magnetic flux enclosed by the interferometers. Thereafter the
system was tuned to a single 2PI. As predicted by theory \cite{Sam04}
no single-particle AB-oscillations in the average current were
observed but the current cross correlations displayed clear
two-particle AB-oscillations, with an amplitude $25 \%$ of the
predicted coherent, zero temperature value. By measuring also the
period of the two-particle oscillations as a function of
interferometer area and enclosed flux and comparing to the sum of the
periods for the two Mach Zehnder interferometers, the two-particle
nature of the AB-oscillations could be established beyond doubt.
\begin{figure}[h]
\centerline{\psfig{figure=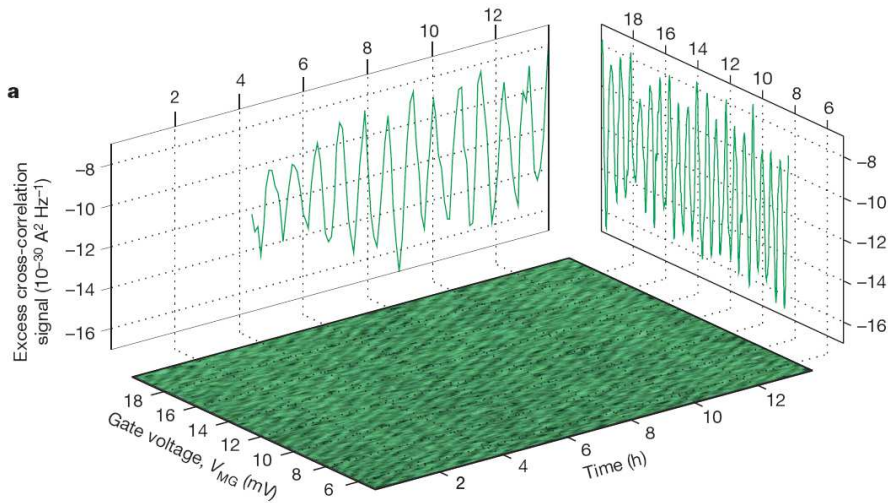,width=12.0cm}}
\caption{Figure reproduced from Ref. \cite{Neder}. Experimental
demonstration of the two-particle AB-effect. Current cross correlation
displaying clear oscillations as a function of the effective
interferometer area and enclosed magnetic flux.}
\label{expfig2}
\end{figure}

In the experiment semitransparent beam splitters were used,
$T_C=T_D=1/2$. For the current cross correlations, theory for finite
temperature and dephasing \cite{Vanessa} predicts, for $A+,B+$,
\begin{equation}
S_{A+B+}=-\frac{e^3V}{4h}H\left[1-\gamma \sin \phi \right].
\label{noise2}
\end{equation}
The temperature dependence is fully contained in
\begin{equation}
H=\coth\left(\frac{eV}{2kT}\right)-\frac{2kT}{eV},
\end{equation}
varying from unity for $kT\ll eV$ to zero for $kT \gg eV$. The effect
of finite temperature is thus to suppress the overall amplitude of the
current cross correlation oscillations. In the experiment, the applied
bias was $7.8\mu V$. The electron temperature was estimated from
independent auto-correlation measurements to be $10mK$. This yields
the temperature suppression factor $H=0.78$. A direct comparison to
Eq. (\ref{noise2}) then gives the oscillation amplitude $H\gamma=0.25$,
i.e. $\gamma=0.32$, a substantial dephasing.

\section{Finite temperature state}

Our main aim of this work is to theoretically investigate the effects
of finite temperature on the entanglement of the state emitted out
from the source, towards the detectors. A prerequisite is to obtain
both a qualitative and a quantitative description of the emitted
many-body state at finite temperature. We consider the experimentally
relevant situation with all source and detector reservoirs kept at the
same temperature $T$. Due to the finite temperature, not only the
electrons emitted from the source in the energy range $0\leq E \leq
eV$ are of interest, we must in principle take into account particles
emitted from all reservoirs at all possible energies. However, due to
the chiral geometry of the 2PI in Fig. \ref{HBTferm}, particles
emitted from the detectors can never scatter back to the detectors,
i.e. detector cross talk is topologically prohibited. The particles
arriving at the detectors thus all originate from the source
reservoirs and we can focus on the many body state emitted by source
$1$ to $4$. We note that in the slightly different geometry realized
experimentally \cite{Neder}, there is the possibility for scattering
between the detectors. It can however be shown \cite{Samnew} that this
does not influence the entanglement of the emitted state.

At finite temperature the state injected from the sources is mixed and
described by a density matrix \cite{Beenrev}
\begin{eqnarray}
\rho_{in}&=&\prod_{E}\rho_{in}(E) \nonumber \\
\rho_{in}(E)&=&\prod_{\kappa=1}^4\left[[1-f_{\kappa}(E)]|0\rangle\langle
0|+f_{\kappa}(E)a^{\dagger}_{\kappa}(E)|0\rangle\langle
0|a_{\kappa}(E)\right]
\label{rhoin}
\end{eqnarray}
where $f_{\kappa}(E)$ is the Fermi distribution of source reservoir
$\kappa=1-4$. The outgoing state is then obtained by inserting the
scattering relations of Eq. (\ref{scatrel}) int Eq. (\ref{rhoin}).

One can see from Eq. (\ref{rhoin}) that the effect of finite
temperature is to give rise to states with 0 to 4 particles emitted at
a given energy. For the terms of interest, i.e. with at least one
particle at both A and B, there is at finite temperature the
possibility for e.g. two particles at A and one at B etc. These terms
are of central importance in the discussion below.

\section{Projected two-particle density matrix}

A theory for entanglement production in non-interacting \cite{Been03}
conductors at finite temperature was presented by Beenakker
\cite{Beenrev} and along similar lines in closed condensed matter
systems by Dowling, Doherty and Wiseman \cite{Wiseman}. At a given
energy, only the component of the emitted many-body state with one
particle in detector region A and one in B has nonzero
entanglement. Moreover, as emphasized in Ref.  \cite{Wiseman}, only
this term describes two particles which each live in a well defined
$2\times 2$ Hilbert spaces at A and B respectively, i.e. two coupled
orbital qubits. We point out that this definition does not take into
account occupation-number, or Fock-space entanglement.  The first step
is thus to project out the two-particle component from the many-body
wave function, which is accomplished by the projection operator
\begin{equation}
\Pi=\Pi_A\otimes\Pi_B, \hspace{0.5cm} \Pi_{\alpha}=n_{\alpha 1}(1-n_{\alpha 2})+n_{\alpha 2}(1-n_{\alpha 1})
\end{equation}
where $n_{Aj}=b_{Aj}^{\dagger}b_{Aj}$ with $j=1,2$ etc is the number
operator (suppressing energy notation). This yields the projected
density matrix
\begin{equation}
\rho_{p}(E)=\Pi\rho(E)\Pi 
\end{equation}
The elements of the density matrix $\rho_{p}(E)$ are conveniently calculated from the relation \cite{Wiseman}
\begin{equation}
[\rho_{p}(E)]_{ij,kl}=\langle \Pi b_{Ai}^{\dagger}b_{Bj}^{\dagger}b_{Bk}b_{Al} \Pi \rangle
\end{equation}
where, for any operator X, $\langle X \rangle=\mbox{tr}[X \rho]$ is
the standard quantum-statistical average. Some algebra gives the
projected density matrix, formally equivalent to the density
matrix calculated in \cite{Beenrev}, Eqs. (B9) - (B13),
\begin{equation}
\rho_p(E)=(1-f)^2f_V^2\left(\begin{array}{cccc} \chi & 0 & 0 & 0 \\ 0 & c_{12}^{12} & c_{12}^{21} & 0 \\ 0 &  c_{21}^{12} & c_{21}^{21} & 0 \\ 0 & 0& 0 & \chi \end{array} \right)
\label{projected}
\end{equation}
where $\chi=e^{-eV/kT}$ and $f$ and $f_V$ the Fermi distribution
functions of the grounded and biased source reservoirs
respectively. The coefficients
\begin{eqnarray}
c_{12}^{12}&=&(R_C[1-\chi]+\chi)(T_D[1-\chi]+\chi), \nonumber \\
c_{21}^{21}&=&(T_C[1-\chi]+\chi)(R_D[1-\chi]+\chi), \nonumber \\
c_{12}^{21}&=&(c_{21}^{12})^*=-\gamma \sqrt{R_CT_CR_DT_D}e^{i\phi_0}(1-\chi)^2
\end{eqnarray}
with $\phi_0$ an overall scattering phase of the beam splitters C and
D. Thus, only the prefactor $f_V^2(1-f)^2$ depends on energy. As for
the zero temperature case we have introduced dephasing as a
suppression of the off-diagonal components of the density matrix. It
follows from Eq. (\ref{projected}) that finite temperature leads to
\\
i) an overall modification of the energy-dependent probability for
two-particle emission via the prefactor $(1-f)^2f_V^2$. \\
ii) a suppression $\sim (1-\chi)^2$ of the off-diagonal components,
equivalent to the effect of dephasing.  \\
iii) a finite amplitude for the diagonal density matrix elements
$[\rho_p(E)]_{11,11}$ and $[\rho_p(E)]_{22,22}$, i.e for two particles
being emitted from either sources 1,3 or 2,4. \\

Additional insight follows from writing the projected density matrix
as
\begin{equation}
\rho_p(E)=(1-f)^2f_V^2\left[\chi
\rho_p^{diag}+(1-\chi)^2\rho^{int}\right]
\label{projected2}
\end{equation}
where the diagonal density matrix 
\begin{equation}
\rho_p^{diag}=\chi \hat 1\otimes \hat 1+(1-\chi)[\rho_A\otimes \hat
1+\hat 1\otimes\rho_B]
\end{equation}
with the zero temperature single particle density matrices
$\rho_A=R_C|1\rangle\langle 1|+R_D|2\rangle\langle 2|$ and
$\rho_B=T_C|1\rangle\langle 1|+T_D|2\rangle\langle 2|$. The density
matrix 
\begin{eqnarray}
\rho^{int}&=&R_CT_D|12\rangle\langle 21|+R_DT_C|21\rangle\langle
12| \nonumber \\
&-&\gamma
\sqrt{T_CR_CT_DR_D}[e^{i\phi_0}|21\rangle\langle 21|+e^{-i\phi_0}|12\rangle\langle
12|]
\end{eqnarray}
results from the two-particle interference. Here we used the shorthand
notation $|12\rangle\equiv|1\rangle_{A}|2\rangle_{B}$ with $\langle
21|=(|12\rangle)^{\dagger}$ etc. Note that the effect of decoherence
enters as a suppression of the two-particle interference
$|\Psi^{int}\rangle\langle \Psi^{int}| \rightarrow \rho^{int}$, where
$|\Psi^{int}\rangle=\sqrt{R_CT_D}|12\rangle-e^{i\phi_0}\sqrt{T_CR_D}|21\rangle$.

Writing $\rho_p(E)$ in the form in Eq. (\ref{projected2}) shows that,
taken the energy dependent prefactor $f_V^2(1-f)^2$ aside, the effects
of finite temperature can be viewed as follows: First, the amplitude
of the two-particle interference component $\rho^{int}$ is suppressed
with increasing temperature as $\sim (1-\chi)^2$. Second, the density
matrix acquires a purely diagonal component $\rho_p^{diag}$ with an
amplitude $\sim \chi$ (note that $\mbox{tr}[\rho_p^{diag}]=4$,
independent on temperature).

\begin{figure}[h]
\centerline{\psfig{figure=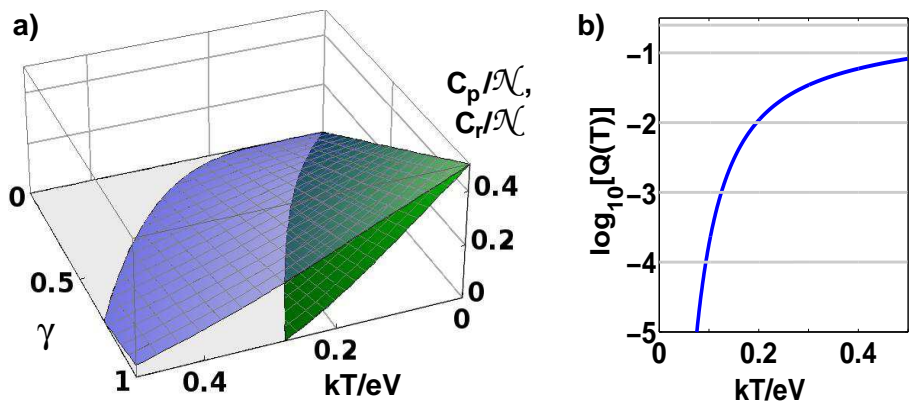,width=12.0cm}}
\caption{a) Entanglement production $C_p/{\mathcal N}$ (blue,
transparent) and $C_r/{\mathcal N}$ (green, opaque) as functions of
temperature $kT/eV$ and coherence $\gamma$ for the semi-transparent
2PI. b) Parameter $Q$ as a function of $kT/eV$ (blue line). Values
$0.25,10^{-1},10^{-2},10^{-3},10^{-4}$ shown (gray lines). Figure reproduced from Ref. \cite{Sam09}.}
\label{fig2}
\end{figure}

For the entanglement, following \cite{Beenrev} we introduce $\sigma_p$
and $w_p(E)$, the normalized density matrix and the emission
probability of the emitted two-particle state respectively, defined
from
\begin{eqnarray}
\rho_p(E)&=&w_p(E)\sigma_p, \nonumber \\
w_p(E)&=&\mbox{tr}[\rho_p(E)]=(1-f)^2f_V^2[(R_CT_D+T_CR_D)(1-\chi)^2+4\chi]
\end{eqnarray}
where we note that $\sigma_p$ is independent on energy. The emission
probability $w_p(E)$ is thus the probability, per unit energy, that
the (normalized) two-particle state $\sigma_p$ is emitted. The
concurrence production per unit energy is then
\begin{eqnarray}
C_p(E)&\equiv& w_p(E)C(\sigma_p)=\frac{(1-\chi)^2f_V^2(1-f)^2}{2} \nonumber \\
&\times& \mbox{max}\left\{4\gamma\sqrt{R_CT_CR_DT_D}-\frac{1}{\sinh^2(eV/2kT)},0\right\}
\label{conceproj}
\end{eqnarray} 
and the total entanglement production during a time $\tau$,
$C_p=(\tau/h) \int dE C_p(E)$, is then (${\mathcal N}=\tau eV/h$)
\begin{equation}
C_p=\frac{{\mathcal N}H}{2}\mbox{max}\left\{4\gamma\sqrt{T_CR_CT_DR_D}-\frac{1}{\sinh^2(eV/2kT)},0\right\}.
\label{concproj}
\end{equation}
We denote this the projected entanglement. As shown in
Fig. \ref{fig2}, $C_p$ decreases monotonically as a function of
$T$. It reaches zero at a critical temperature $T_c^p$ given by
\begin{equation}
kT_c^p=eV \ln \left(\frac{\sqrt{1+4\gamma\sqrt{R_CT_CR_DT_D}}+1}{\sqrt{1+4\gamma \sqrt{R_CT_CR_DT_D}}-1}\right)
\label{critTc}
\end{equation}
For semi-transparent beam-splitters and zero dephasing, $\gamma=1$,
the entanglement thus survives up to \cite{Beenrev} $kT_c^p=0.57 eV$.

Inserting the parameter values from the experiment, we get $C_p
\approx 0.1{\mathcal N}$ and $C(\sigma_p)\approx 0.3$, i.e. {\it the
state emitted by the 2PI is clearly entangled}. Importantly, the
effect of finite temperature is essentially negligible, the reduction
in entanglement comes from decoherence.

The entanglement of the projected density matrix is the entanglement
one could access, had one been able to do arbitrary local operations
and classical communication between A and B, i.e. fully energy and
particle resolved measurements. Under realistic conditions this is not
possible, the accessible physical quantities are currents and current
cross correlators. Is it possible to determine the projected
entanglement with such measurements? The answer to this question is
no, for two main reasons: \\ i) As discussed above, at nonzero
temperatures it is not only the biased source reservoirs which emit
particles but also the grounded source reservoirs do. As a
consequence, there is a finite amplitude for emitted states with
two-particles at A and/or at B. These unentangled states contribute to
currents and current correlators, which results in a detectable state
with suppressed entanglement. \\ ii) The current and current
correlators provide information on the energy integrated properties of
the many-body state, not on the emitted state at each energy. This
lack of energy-resolved information leads to a further suppression of
the detectable entanglement. \\ Clearly, these effects of the
thermally excited Fermi sea constitute generic problems when trying to
detect entanglement in mesoscopic conductors.

As a remedy for these finite temperature read-out problems it was
suggested to work with detectors at very low temperatures
\cite{Beenrev}. Another idea was recently presented by Hannes and
Titov \cite{Titov}. They investigated detection of entanglement at
finite temperatures via a Bell inequality and proposed to introduce
energy filters at the drains. However, both schemes
\cite{Beenrev,Titov} would lead to additional experimental
complications in systems which already are experimentally very
challenging. Our idea is instead to investigate what information about
the projected entanglement can actually be deduced from current and
current correlation measurements.

In this context we also mention the recent proposal by Kindermann
\cite{Kindermann}, to produce and detect entangled electron-hole pairs
in graphene via a Bell inequality formulated in terms of the transport
part of the current cross correlators \cite{mb92}, i.e. by subtracting
away the thermal equilibrium correlators from the finite bias ones. In
our work \cite{Sam09} we proposed a similar scheme for a general
mesoscopic conductor. However, as was pointed out in \cite{Sam09} and
is further discussed below, it is important that one performs a
detailed comparison of the projected entanglement and the entanglement
obtained from current cross correlation measurements. Without such a
comparison, there is the possibility that one concludes, based on
correlation measurements, finite entanglement where there is none,
i.e. the projected entanglement is zero.

\section{Reduced two-particle density matrix}

We first consider the expression for the current and zero frequency
current cross correlators at contacts $A+$ and $B+$ at finite
temperatures. We have
\begin{eqnarray}
I_{A+}&=&\frac{e}{h}\int dE \left[\langle n_{A+}\rangle-f\right],
\hspace{0.5cm} I_{B+}=\frac{e}{h}\int dE \left[\langle
n_{B+}\rangle-f\right], \nonumber \\ S_{A+B+}&=&\frac{e^2}{h}\int dE
\langle \Delta n_{A+} \Delta n_{B+}\rangle
\label{currnoisered}
\end{eqnarray}
where $\langle \Delta n_{A+} \Delta n_{B+} \rangle=\langle
n_{A+}n_{B+}\rangle-\langle n_{A+}\rangle \langle n_{B+} \rangle$ is
the irreducible correlator. As discussed above, the many-body state
incident on the detectors originates from the sources. It is the
properties of this state that determines the observables $\langle
n_{A+}\rangle,\langle n_{B+}\rangle$ and $\langle \Delta n_{A+} \Delta
n_{B+}\rangle$ and thus establishes a connection between the emitted
state and the physical quantities accessible in a measurement.

\subsection{Energy resolved reduced density matrix}

In order to better understand the readout problem discussed above, we
first discuss the energy resolved properties of the emitted state. If
one would have access to energy filters, as proposed in \cite{Titov},
or would be working at zero temperature, by combining current and
current cross correlations it would be possible to get direct access
to the energy resolved quantities $\langle n_{A+}\rangle,\langle
n_{B+}\rangle$ and $\langle \Delta n_{A+} \Delta n_{B+}\rangle$. As is
discussed below, by a suitable set of measurements with different
settings of the beam splitters at A and B one could then
tomographically reconstruct the (unnormalized) density matrix of the
state emitted out from the source beam splitters C and D, $\rho_r^E$,
with elements given by
\begin{equation}
[\rho_r^E]_{ij,kl}=\langle b_{Ai}^{\dagger}b_{Bj}^{\dagger}b_{Bk}b_{Al} \rangle
\end{equation}
We denote $\rho_r^E$ the energy resolved reduced density matrix.

By comparing $\rho_r^E$ with the expression for the projected density
matrix in Eq. (\ref{projected}) we see that it differs by the
projection operators. Consequently, the reduced density matrix
contains also the contributions from processes with more than one
particle at A and/or at B. After some algebra we find the density
matrix
\begin{equation}
\rho_r^E=(1-f)^2f_V^2\left(\begin{array}{cccc} \tilde\chi & 0 & 0 & 0
\\ 0 & \tilde c_{12}^{12} & c_{12}^{21} & 0 \\ 0 & 
c_{21}^{12} & \tilde c_{21}^{21} & 0 \\ 0 & 0& 0 & \tilde \chi
 \end{array} \right)
\label{redendens}
\end{equation}
where we introduced $\tilde \chi=\chi/[(1-f_V)(1-f)]$ and the coefficients
\begin{eqnarray}
\tilde c_{12}^{12}&=&(R_C[1-\chi]+\chi)(T_D[1-\chi]+\tilde \chi), \nonumber \\
\tilde c_{21}^{21}&=&(T_C[1-\chi]+\chi)(R_D[1-\chi]+\tilde \chi).
\end{eqnarray}
A comparison to the projected density matrix in Eq. (\ref{projected})
shows that $\rho_r^E$ only differs formally from $\rho_p(E)$ by the
change $\chi \rightarrow \tilde \chi$ at a number of places. This has
the consequence that the normalized density matrix
$\sigma_r^E=\rho_r^E/w_r^E$, with $w_R^E=\mbox{tr}[\rho_r^E]$ depend
on energy. That is, in contrast to $\rho_p$ both the normalized,
emitted two-particle state as well as the emission probability depend
on energy. Qualitatively, as discussed above, the difference between
$\rho_r^E$ and $\rho_p(E)$ arises from the fact that also states with
more than one particle at A and/or B contribute to $\rho_r^E$ but not
to $\rho_p(E)$. Writing $\rho_r^E$ on a form similar to
Eq. (\ref{projected2}) one sees that these three and four particle
states contribute only to the diagonal part of $\rho_r^E$.

\begin{figure}[h]
\centerline{\psfig{figure=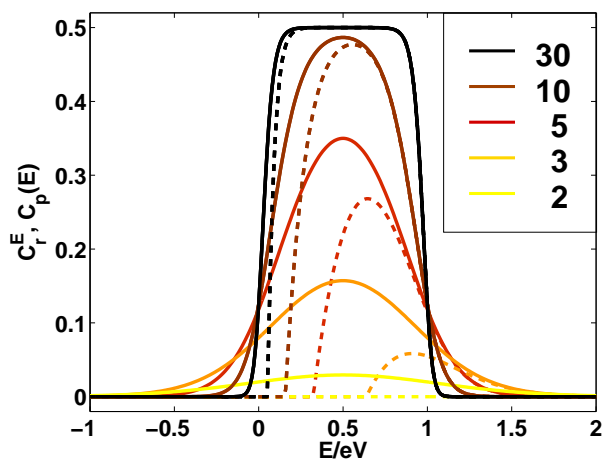,width=8.0cm}}
\caption{A comparison of the concurrence production rates $C_r^E$
(dashed) and $C_p(E)$ (solid), as a function of energy for
$T_C=T_D=1/2$ and different ratios $eV/kT$.}
\label{redeconc}
\end{figure}

Turning to the entanglement, the concurrence production
$C_r^E=w_r^EC(\sigma_r^E)$ at energy $E$ is then
\begin{eqnarray}
C_r^E&=&\frac{(1-\chi)^2f_V^2(1-f)^2}{2} \nonumber \\
&\times&
\mbox{max}\left\{4\gamma\sqrt{R_CT_CR_DT_D}-\frac{1}{\sinh^{2}(eV/2kT)}\frac{1}{(1-f_V)(1-f)},0\right\}
\label{concered}
\end{eqnarray}
From the expression for the concurrence it becomes clear that the
separable three and four-particle states are detrimental for the
entanglement. Hence, finite temperature leads to a stronger
suppression of the reduced, energy resolved density matrix than of the
projected one. This is illustrated in fig. \ref{redeconc} where the
corresponding concurrencies are plotted for semitransparent
beam-splitters and different values of $kT/eV$.
As is clear from the figure, there is an energy $E_0$ above which the
concurrence is finite (up to $E \rightarrow \infty$). The energy $E_0$
is given by the condition $C_r^E(E_0)=0$, as
\begin{equation}
E_0=kT \left(\ln[2]-\ln \left[(1-\chi)\sqrt{1+4\sqrt{R_CT_CR_DT_D}}-(1+\chi)\right]\right)
\label{encond}
\end{equation}
What is moreover clear from Fig. \ref{redeconc} is that, for all
energies, $C_r^E(E)<C_p(E)$. The difference is obvious for energies
$E<E_0$, where $C_r^E=0$. At these energies the probability for
emission of separable three and four particle states is thus large
enough to completely suppress the entanglement of the reduced density
matrix.

Importantly, the relation $C_r^E(E)<C_p(E)$ holds for all settings of
the beam splitters $T_C$ and $T_D$, as is clear by comparing
Eqs. (\ref{conceproj}) and (\ref{concered}). The reason for this is
that the reduced density matrix contains contributions from all
individual particle density matrices $\sigma_{ij}$ with $i,j\geq 1$
(e.g. $\sigma_{12}$ describes one particle at A and two at B) while
the projected density matrix only depends on $\sigma_{11}$. Since all
$\sigma_{12},\sigma_{21},\sigma_{22}$ are separable and the
concurrence is a convex quantity, i.e. $C(p_1\sigma_1+p_2\sigma_2)\leq
p_1 C(\sigma_1)+p_2 C(\sigma_2)$ for $p_1+p_2=1$, the concurrence
$C_r^E$ is always smaller than $C_p(E)$. We point out that this
carries over to the total concurrence production found by integrating
Eq. (\ref{concered}) over energy (result not presented here).

It follows from Eq. (\ref{encond}) that for a critical temperature
$T_c^{rE}$ the energy $E_0 \rightarrow \infty$, i.e. the entanglement is
zero for any energy. Interestingly, this happens for the same
temperature as for the projected concurrence, Eq. (\ref{critTc}).

\subsection{Finite temperature reduced density matrix}

Importantly, at finite temperature, without any energy filters, we do
not have access to the energy resolved quantities discussed above,
only to the total currents and current correlators measured at
contacts $A\alpha,B\beta$. In Ref. \cite{tomo} it was discussed how
to, at zero temperature, tomographically reconstruct the reduced
density matrix using currents and current correlations. Extending this
scheme to nonzero temperatures it is natural to define the finite
temperature reduced density matrix $\rho_r$ via the relation
\begin{eqnarray}
&&\frac{I_{A\alpha}I_{B\beta}}{(Ve^2/h)^2}+\frac{S_{A\alpha
B\beta}}{2Ve^3/h}= \mbox{tr}\left\{\left[I_{A\alpha}^O\otimes
I_{B\beta}^O \right]\rho_r\right\}.
\label{noisecurrrel}
\end{eqnarray}
We emphasize that $\rho_r$ is reconstructed from observables already
integrated over energy and does hence not depend on energy. Also note
that $\rho_r$ is not given by integrating $\rho_r^E$ over energy, in
fact the difference between the two density matrices is further
discussed below.

In Eq. \ref{noisecurrrel} the orbital current operators in the local
basis $\{|1 \rangle, |2\rangle \}$, including the rotations at the
detector splitters, are $I_{A\alpha}^O=(\hat 1+\alpha {\mathbf
n}_A\cdot \hat \sigma)/2$ and $I_{B\beta}^O=(\hat 1+\beta {\mathbf
n}_B\cdot \hat \sigma)/2$, with ${\mathbf n}_A\cdot \hat
\sigma=S_A\sigma_zS_A^{\dagger}$ and ${\mathbf n}_B\cdot \hat
\sigma=S_B\sigma_zS_B^{\dagger}$ where $\hat
\sigma=[\sigma_x,\sigma_y,\sigma_z]$ a vector of Pauli matrices and
$S_A~(S_B)$ the scattering matrix of the beam splitter at A (B).

Making use of the results for finite temperature current and current
correlations in \cite{Vanessa} we obtain the reduced density matrix
\begin{equation}
\rho_r=\left(\begin{array}{cccc} R_CT_C(1-H) & 0 & 0 & 0 \\ 0 & R_CT_D
& d^{12}_{21} & 0 \\ 0 & d_{12}^{21} & R_DT_C & 0 \\ 0 & 0& 0 & R_DT_D(1-H)
\end{array} \right)
\label{reduced}
\end{equation}
where $d^{12}_{21}=(d_{12}^{21})^*=-H\gamma \sqrt{R_CT_CR_DT_D}
e^{i\phi_0}$. Comparing $\rho_r$ to both $\rho_p(E)$ and $\rho_r^E$ in
Eqs. (\ref{projected}) and (\ref{redendens}) it is clear that the
qualitative effect of finite temperature is the same for the reduced
density matrix. The quantitative effects are however different. First,
the temperature dependence enters via $H$ rather than via $\chi$,
giving a much stronger effect of finite temperature. This is the
effect of having access to energy integrated quantities only. Second,
in the expression for the average current in Eq. (\ref{currnoisered}),
in the integrand one subtracts $f$ which arises due to particles
flowing out of the detector reservoirs. This yields smaller diagonal
terms, to be further discussed below.

It is illuminating, just as for $\rho_p(E)$, to write $\rho_r$ as a
sum of a diagonal and an interference part,
\begin{equation}
\rho_r=(1-H)[\rho_A\otimes\rho_B]+H\rho^{int}.
\label{reduced2}
\end{equation}
From this we see that the effect of increasing temperature is to
monotonically increase the amplitude for the separable product state
$\rho_A \otimes \rho_B$, while the amplitude of the interference
component is suppressed. We can thus conclude the following properties
for all three density matrices $\rho_p(E), \rho_r^E$ and $\rho_r$: \\
i) At zero temperature they all reduce to the same expression,
$\rho^{int}$. \\ ii) Increasing temperature leads to a monotonic
suppression of the two-particle interference component. \\ iii) Finite
temperature introduces an additional diagonal component, different for
the three density matrices.
 
Turning to entanglement, introducing the normalized reduced density
matrix $\sigma_r$ we can write
\begin{eqnarray}
\rho_r&=&w_r\sigma_r \nonumber \\
w_r&=&\mbox{tr}[\rho_r]=[R_CT_C+R_DT_D](1-H)+R_CT_D+R_DT_C.
\end{eqnarray}
We then define the
total entanglement production during a time $\tau$ as $C_{r}\equiv
{\mathcal N} w_rC(\sigma_r)$. It is
\begin{equation}
C_r=2{\mathcal N}\mbox{max}\{\sqrt{T_CR_CT_DR_D}[H(1+\gamma)-1],0\}
\label{concred}
\end{equation}
here called the reduced entanglement. As $C_p$, $C_r$ decreases
monotonically with increasing $T$. It reaches zero at a critical
temperature $T_c^r$ given by the relation
\begin{equation}
H(T_c^r)=\frac{1}{1+\gamma}
\end{equation}
For perfect coherence, $\gamma=1$, we have $kT_c^r=0.28eV$, close to
one half of $kT_c^p$. Importantly, in contrast to $T_c^p$, $T_c^r$ is
independent on the setting of the beam splitters.

By comparing the expressions for the two quantities of main interest,
the projected and reduced concurrencies, $C_p$ in Eq. (\ref{concproj})
and $C_r$ in Eq. (\ref{concred}), we can conclude the following: \\ i)
For both $C_p$ and $C_r$ the origin of the entanglement is the
two-particle interference, in fact the component $\rho^{int}$ gives
rise to the positive term $2{\mathcal N}H\gamma\sqrt{T_CR_CT_DR_D}$,
identical for $C_p$ and $C_r$. \\ ii) For both $C_p$ and $C_r$ finite
temperature introduces a negative term, $-{\mathcal
N}H/[2\sinh^2(eV/2kT)]$ for $C_p$ and $-2{\mathcal
N}(1-H)\sqrt{T_CR_CT_DR_D}$ for $C_r$, which leads to a suppression of
the concurrence. These terms arise from the separable, diagonal
components of the corresponding density matrices.

\section{Entanglement bound}

Comparing Eqs. (\ref{concproj}) and (\ref{concred}) quantitatively we
find that $C_p\geq C_r$ for
\begin{equation}
Q(T)=\frac{H}{4(1-H)\sinh^2(eV/2kT)}\leq \sqrt{T_CR_CT_DR_D},
\label{bondcond}
\end{equation}
independent on $\gamma$ (see Fig. \ref{fig2}). Consequently, for beam
splitters away from the strongly asymmetrical (tunneling) limit, {\it
the reduced entanglement constitutes a lower bound for the projected
entanglement}. In the tunneling limit, however, the reduced
entanglement is larger than the projected one. Thus, in contrast to
the energy-resolved reduced density matrix $\rho_r^E$, $\rho_r$ can be
more entangled than $\rho_p$. The origin of this difference is, as
pointed out above, that when calculating (and measuring) $\rho_r$ the
average currents flowing out from the detector reservoirs are
subtracted, yielding a smaller diagonal component and hence a larger
entanglement $C_r$. Importantly, since the transparencies $T_C$ and
$T_D$ can be controlled and measured via average currents in the
experiment, it is always possible to verify independently that the
condition in Eq. (\ref{bondcond}) is satisfied.

Turning to the experiment \cite{Neder}, for the relevant parameters we
have $Q(T)\approx 4\times 10^{-4} \ll \sqrt{R_CT_CR_DT_D}\approx
0.25$, showing the validity of the bound. However, $C_r\approx
0.01{\mathcal N}$ and based on the measurement \cite{Neder} no
conclusive statement can be made about $C_r$ and hence not about
$C_p$. In order to detect entanglement via measurements of currents
and current correlations, one thus need to work at even lower
temperature and further reduce the dephasing in the experiment.

A more detailed understanding of this finite temperature readout
problem can be obtained by comparing the properties of $\sigma_p$ and
$\sigma_r$. For perfect coherence $\gamma=1$ and identical beam
splitters $T_C=T_D={\mathcal T}=1-{\mathcal R}$ one can (up to a local
phase rotation) write 
\begin{equation}
\sigma_{p/r}=\frac{1}{4}\xi_{p/r}\hat 1\otimes
\hat 1+(1-\xi_{p/r})|\Psi_s\rangle\langle \Psi_s|
\end{equation}
a Werner state \cite{Werner}, with singlet weight [$|\Psi_s\rangle$ is
the singlet in Eq. (\ref{introsing})]
\begin{equation}
1-\xi_p=\frac{2\mathcal{RT}\sinh^2(2eV/kT)}{1+2\mathcal{RT}\sinh^2(2eV/kT},
\hspace{0.5cm} 1-\xi_r=\frac{H}{2-H}
\label{singweights}
\end{equation}
Increasing $kT/eV$ from zero, $\xi_p\approx
2e^{-4eV/kT}/(\mathcal{RT})$ becomes exponentially small while $\xi_r
\approx kT/eV$ increases linearly. These qualitatively different
behaviors, clearly illustrated in Fig. \ref{fig2}, are a striking
signature of how a small $kT/eV$, having negligible effect on
$C(\sigma_p)$, leads to a large suppression of $C(\sigma_r)$.

From Eqs. (\ref{concproj}) and (\ref{concred}) follows also a
counter-intuitive result: {\it finite amplitude of the AB-oscillations
is no guarantee for finite two-particle entanglement}. This is
apparent for $\sigma_r$ in the limit of no decoherence $\gamma=1$ and
identical beam splitters $T_C=T_D$, since a separable Werner state,
$\xi_r>2/3$, can be decomposed \cite{decomp} as
\begin{equation}
\sigma_r=\frac{1}{4}\sum_{n=1}^4|\phi^A_n\rangle\langle\phi_n^A|\otimes|\phi_n^B\rangle\langle\phi_n^B|
\label{sepstate}
\end{equation}
with the normalized states at $A$ and $B$
\begin{eqnarray}
|\phi^{A/B}_n\rangle&=&\cos
\theta^{A/B}_n|1\rangle+e^{i\pi[1-2n]/4}\sin \theta^{A/B}_n|2\rangle, \nonumber \\
\theta_1^{A/B}&=&\theta_3^{A/B}=\mbox{atan}[y^{A/B}], \hspace{0.5cm}
\theta_2^{A/B}=\theta_4^{A/B}=-\mbox{acot}[y^{A/B}] \nonumber \\
 y^{A/B}&=&\frac{\sqrt{2-\xi_r}+\sqrt{3\xi_r-2}}{\sqrt{\xi_r}\pm
\sqrt{4-3\xi_r}}, \hspace{0.5cm} +(-)~ \mbox{for}~ A(B)
\end{eqnarray}
This {\it classically} correlated state gives, via
Eq. (\ref{noisecurrrel}), AB-oscillations with amplitude
$2(1-\xi_r)/(2-\xi_r)=H$. Moreover, the reduced local single particle
states are completely featureless,
$\mbox{tr}_B(\sigma_r)=\mbox{tr}_A(\sigma_r)=\hat 1/2$ which means
that there is no single particle Aharonov-Bohm effect. The existence
of classically correlated two-particle states giving rise to
Aharonov-Bohm oscillations in the current cross correlations but not
in the currents provides further motivation for a complete tomographic
reconstruction of the reduced density matrix in order to provide an
unambiguous experimental demonstration of entanglement.

\section{Detecting entanglement: Quantum State Tomography and Bell Inequality}

\subsection{Quantum state tomography}
As pointed out at several places above, the reduced density matrix can
be reconstructed by a suitable set of current and current correlations
measurements with different settings of the beam splitters parameters,
i.e. different ${\mathbf n}_A,{\mathbf n}_B$. A detailed description
of this scheme is given in \cite{tomo}. Here we only emphasize that
the necessary tools, controllable reflectionless electronic beams
splitters and phase gates, are experimentally available, as
demonstrated in e.g. \cite{MZ1,MZ2,MZ3,MZ4,MZ5,Neder}

\subsection{Bell Inequality}
Another widely discussed \cite{BI1,Sam03,Been03,Sam04,nonint1,nonint2}
approach to detect the entanglement in mesoscopic conductors is to
use a Bell inequality. Violation of a CHSH-Bell inequality \cite{CHSH}
formulated in terms of currents and low-frequency current correlations
demonstrates finite entanglement of $\rho_r$. We point out that an
optimal Bell test, requiring control over all three components of
${\bf n_A}$ and ${\bf n_B}$, demands the same number of measurement
and level of experimental complexity as a tomographic reconstruction
of $\rho_r$. The CHSH-Bell inequality is
\begin{equation}
\Omega_{Bp/r}\leq 2
\end{equation}
where $\Omega_{Bp/r}$ is the Bell parameter for the projected/reduced
state. The Bell parameter is formally determined by the
projected/reduced density matrix $\sigma_{p/r}$ and different settings
of the detector beam splitters, reaching its maximum value
$\Omega_{Bp/r}^{max}$ for an optimal setting of ${\bf n_A}$ and ${\bf
n_B}$. From $\sigma_p$ and $\sigma_r$ above, we can, using Ref.
\cite{Horodecki}, calculate the maximal Bell parameters. For symmetric
beam splitters, $T_C=T_D={\mathcal T}$, we have the simple result
\begin{eqnarray}
\Omega_{Bp/r}^{max}&=&2\sqrt{1+\gamma^2}(1-\xi_{p/r})
\end{eqnarray}
where the singlet weights $1-\xi_p$ and $1-\xi_r$ are given in
Eq. (\ref{singweights}). This shows that the effects of decoherence
and finite temperature enters separately in the Bell
parameter. Moreover, as pointed out in
Refs. \cite{Sam03,Turkbeen,Turksam}, at zero temperature a Bell
inequality can in principle be violated for arbitrary dephasing. We
also point out that a detailed investigation of conditions for
violation of a Bell inequality in the presence of dephasing, in the
solid state, was recently performed in Ref. \cite{Kofman}.

The limiting value for violation $\Omega_{Bp/r}^{max}=2$ for
${\mathcal T}=1/2$ plotted in Fig. \ref{fig2}. It is clear that for
the values $kT/eV$ and $\gamma$ of the 2PI-experiment, while
$\Omega_{Bp} \leq 2$ in principle can be violated, a detection of
entanglement by violating $\Omega_{Br} \leq 2$ is not possible. This
demonstrates in a striking way the known fact \cite{Werner,Verstrate}
that there are entangled states that do not give a violation of a Bell
Inequality.

\section{Conclusions}
In conclusion, we have investigate the effect of finite temperature on
the entanglement production and detection in the fermionic
two-particle interferometer, presenting an extended discussion of the
results in Ref. \cite{Sam09}. A calculation of the entanglement of the
two-particle state projected out from the emitted, finite temperature
many body state shows that the state emitted in the two-particle
interferometer in the experiment by Neder et al \cite{Neder} is
clearly entangled. By comparing the entanglement of the projected
two-particle state with the entanglement of the reduced two-particle
state, accessible via quantum state tomography based on current and
current correlation measurements, we establish that the entanglement
of the reduced state constitute a lower bound for the entanglement of
the projected state. In the two-particle interferometer experiment the
reduced state is however marginally entangled. Moreover, a finite
temperature Bell Inequality formulated in terms of currents and
current correlators can not be violated in the experiment. This shows
that an unambiguous demonstration of the entanglement via measurements
of currents and current correlations requires a reduction of the
dephasing and the temperature.

\section{Acknowledgements}

The work was supported by the Swedish VR, the Israeli SF, the MINERVA
foundation, the German Israeli Foundation (GIF) and Project
Cooperation (DIP), the US-Israel Binational SF, the Swiss NSF and
MaNEP.

\end{document}